\documentclass[prb,aps,twocolumn,showpacs]{revtex4}

\usepackage{graphicx}

\usepackage{amsmath,bm,amsfonts}
\newcommand{\kk}{\bm{k}}

\newcommand{\bk}{b_{\bm{k}}}
\newcommand{\bkd}{b_{\bm{k}}^{\dag}}

\newcommand{\sumk}{\sum_{\bm{k}}}

\newcommand{\rl}{\rangle\!\langle}
\newcommand{\mev}{\mathrm{meV}}

\begin{document}

\author{Piotr Kaczmarkiewicz}
 \email{piotr.kaczmarkiewicz@pwr.wroc.pl}
\author{Pawe{\l} Machnikowski}
 \email{pawel.machnikowski@pwr.wroc.pl}
 \affiliation{Institute of Physics, Wroc{\l}aw University of
Technology, 50-370 Wroc{\l}aw, Poland}

\title{Two-phonon polaron resonances in self-assembled quantum dots}

\begin{abstract}
We study the second-order polaronic resonance between 2-LO-phonon
states and p-shell electron states in a quantum dot. We show that the
spectrum in the resonance area can be quantitatively reproduced by a
theoretical model. We propose also a perturbative approach to the
problem based on a quasi-degenerate perturbation theory. This method
not only considerably reduces the numerical complexity without
considerable loss of accuracy but also gives some insight into the
structure and origin of the resonance spectrum.
\end{abstract}


\maketitle

\section{Introduction}

Carrier-phonon interaction is one of the major factors that determine
the optical and transport properties of semiconductors. In
semiconductor quantum dots, phonon-related effects manifest themselves
in a special way due to the discrete spectrum of carriers confined in
these structures. On the one hand, such a discrete density of states
may slow down carrier relaxation
\cite{bockelmann90,urayama01,heitz01}.  On the other hand, coupling of
a confined charge system to the nearly dispersionless (thus, also
spectrally discrete) system of longitudinal optical (LO) phonons leads
to the formation of correlated carrier-phonon states
\cite{hameau99,verzelen02a} and to a reconstruction of the system
spectrum, as observed, e.g., in intraband absorption experiments
\cite{hameau99,hameau02}. This effect is a generalized form of a
polaron, known from bulk systems.

Apart from the (unmeasurable) energy shifts accompanying the polaron
formation, this effect manifests itself by the appearance of
polariton-like resonances (anticrossings) between the essentially
discrete zero-phonon and 1- or 2- phonon states whenever the
energy difference between two carrier states matches the energy of one
or two LO phonons \cite{hameau02}. The 2-phonon resonance is of
particular interest since the 2-LO-phonon energy of about 70~meV falls
into the range of typical separations between electron energy levels
in self-assembled structures.

Understanding such coherent effects in the carrier-phonon coupling
in quantum dots is essential not only for the correct description of
the system spectrum but also for the discussion of relaxation
properties. In particular, the strong coupling between the confined
carriers and LO phonons precludes purely LO-phonon-mediated relaxation
but opens new channels of efficient two-phonon (acoustic + optical)
emission \cite{verzelen00,jacak02a,grange07}, which may explain
efficient carrier relaxation observed in some experiments
\cite{heitz97,ignatiev01,zibik04}.

Far from the degeneracy (resonance) point, the coupling to LO phonons
may be treated perturbatively. Also a first-order polaronic resonance
does not present serious difficulties, as it involves only a
zero-phonon state and a one-phonon state. These two states are
directly coupled by the Fr\"ohlich interaction Hamiltonian 
and the resulting resonant anticrossing can be treated, e.g., by a
second-order Wigner--Brillouin perturbation theory 
\cite{wendler93a,wendler93b}. In contrast, the
second-order resonance, involving a zero-phonon state and a two-phonon
state is more complex, since the observed anticrossing is due to
higher-order, indirect couplings via other states. In this case one
resorts to a numerical treatment
\cite{hameau99,hameau02,jacak03b,obreschkow07,stauber00}. The latter
is feasible due to the discrete (dispersionless) character of the LO
modes, which allows one to describe the LO phonon subsystem by a
finite set of either orthogonal \cite{stauber00} or non-orthogonal
\cite{obreschkow07} collective modes. The numerical approach to the
polaron problem turns out to successfully reproduce the qualitative
features. However, the early attempts to model
the experimental results \cite{hameau99,hameau02} suggested
that certain material parameters must be adjusted in order to achieve
a quantitative agreement with the measurement data. This seemed
plausible\cite{jacak02f} as QDs are inherently inhomogeneous systems
the quantitative 
properties of which may depend, e.g., on the size and composition
characteristics. Such a viewpoint would mean, however, that a full
understanding and quantitative description of the carrier--LO phonon
interaction in QDs may not be possible.

The goal of the present paper is twofold. First, we discuss the
spectrum of the electron-LO-phonon system in the vicinity of the
two-phonon resonance, including the effects of QD asymmetry
(ellipticity), non-parabolicity of the confinement potential, as well
as the external magnetic field. We show that the experimentally
observed spectrum of the coupled electron-LO-phonon system can be
successfully reproduced using only standard material constants,
without any adjustable parameters. In this way we show that the
existing theory\cite{stauber00,hameau02} does allow us
to understand the physics of carrier--LO phonon interactions in QDs
completely and quantitatively. 
We calculate also the intraband (far infra-red)
absorption spectrum of a QD ensemble, where the polaronic resonance is
clearly manifested. Second, we present an effective Hamiltonian
approach, based on the quasi-degenerate perturbation theory which
reduces the problem size from a few thousands of basis states to just
a few and provides some insight into
the structure of two-phonon polaron states. This, in turn,
demonstrates that describing the electron--LO phonon system does not
necessarily have to involve heavy numerics and may open the way to
the efficient description of even higher-order effects.

The paper is organized as follows: First, in Sec. \ref{sec:model}, we
recall the model of an electron confined in a QD interacting with LO
phonons.  Then in Sec. \ref{sec:numerics}, numerical approach is
described.  The spectrum of a single QD, as well as absorption on a QD
ensemble and the magnetopolaron spectrum are presented in
Sec. \ref{sec:numres}. The two following sections \ref{sec:Heff} and
\ref{sec:resHeff} describe the effective Hamiltonian approach and
present results obtained within this method. 

\section{The model}
\label{sec:model}

We consider a self-assembled quantum dot occupied by a single
electron. The effective confinement potential in the $xy$ plane is
assumed to be almost axially symmetric and parabolic, although small
corrections accounting for ellipticity and non-parabolicity will be
taken into account. In the $z$ (growth) direction, the confinement is
assumed to be much stronger, as is typical for these structures. The
electron confined in the QD is coupled to the polarization field
associated with the LO phonons. In calculations, GaAs parameters will
be used.

The system is described by the Hamiltonian
\begin{eqnarray}\label{ham}
H=H_{0}+H_{\mathrm{a}}+H_{\mathrm{np}}+H_{\mathrm{int}}+H_{\mathrm{ph}},
\end{eqnarray} where $H_{0}$ describes an electron in an isotropic
dot, $H_{\mathrm{a}}$ and $H_{\mathrm{np}}$ account for the anisotropy
(ellipticity) and non-parabolicity of the confinement potential,
respectively, $H_{\mathrm{int}}$ describes the electron-phonon
coupling and $H_{\mathrm{ph}}$ is the free LO phonon Hamiltonian.

The first term in Eq.~(\ref{ham}) is \cite{jacak98a}
\begin{equation*} H_{0}=\frac{1}{2m^{*}}\left ( {\bm{p}} - e
{\bm{A}}\right )^2 + \frac{1}{2}m^{*}\omega_0^2 r_{\bot}^2 +
\frac{1}{2} m^{*} \omega_z^2 z^2
\end{equation*} 
and describes an electron in an axially symmetric
harmonic potential in a magnetic field $B$ oriented along the symmetry
axis, where $m^{*}=0.066m_{\mathrm{e}}$ is the effective mass of an
electron in GaAs and $\bm{r_{\bot}}$ denotes the in-plane component of
the electron position. The energy $\hbar \omega_0$ is the level
spacing for in-plane excitations in the 
absence of magnetic fields.  We will refer to this quantity as the characteristic
energy of the system and use it as a system parameter in the
discussion that follows.  
This energy parameter is related to the in-plane confinement length
$l_0 = \sqrt{\hbar/(m^*\omega_0)}$.
The confinement along the $z$ axis is much
stronger than that in the $xy$ plane, that is, $\omega_{z}\gg
\omega_{0}$.  The dynamics along this strongly confined direction is
restricted to the lowest subband, corresponding to the ground state
wave function
\begin{equation*} \Phi_{z}(z)=\frac{1}{\sqrt{l_z} {\pi}^{1/4}}
e^{-\frac{z^2}{2l_z^2}},
\end{equation*} where $l_{z}=\sqrt{\hbar/(m^{*}\omega_{z})}$ is the
confinement length in this direction. 
Since the dots we intend to model have the height to diameter ratio of
about 10 (Ref.~\onlinecite{hameau02}) we choose $l_z=0.1 l_0$, which
will be fixed throughout the paper.

The essential part of $H_{0}$, accounting for the dynamics in the $xy$
plane, is the well known Fock--Darwin Hamiltonian describing a
2-dimensional harmonic oscillator in a perpendicular magnetic field.
We choose the symmetric gauge, ${\bm{A}}=\frac{1}{2}(-By, Bx, 0)$,
where B is the magnetic field, and use the basis of eigenstates of this
Hamiltonian (Fock--Darwin states \cite{jacak98a}), denoted as
$|nm\rangle$, where $n=0,1,\ldots$ and $m=\ldots,-1,0,1,\ldots$ are
the radial and angular momentum quantum numbers, respectively (that
is, $\hbar m$ is the projection of the angular momentum on the
symmetry axis $z$).  The corresponding wave functions are
\begin{eqnarray} 
\lefteqn{\Psi_{nm}(\bm{r_{\bot}})=\langle
\bm{r_{\bot}}|nm\rangle =}\nonumber\\ 
&&\frac{\sqrt{2}}{l_B} \sqrt{\frac{n
!}{(n + |m|)!}} \left ( \frac{r_{\bot}}{l_B}\right )^{|m|}
e^{-\frac{r_{\bot}^2}{2l_B^2}} \mathcal{L}^{|m|}_{n} \left (
\frac{r_{\bot}^2}{l_B^2} \right )\nonumber,
\end{eqnarray} where $\mathcal{L}^{|m|}_{n} \left ( s \right )$ is
a Laguerre polynomial.  Here
$l_{\mathrm{B}}=\sqrt{\hbar/(m^{*}\omega_{\mathrm{B}})}$ is the
in-plane confinement width in the magnetic field, where
$\omega_{\mathrm{B}}^{2}=\omega_{0}^{2}+\omega_{\mathrm{c}}^{2}/4$
and $\omega_{\mathrm{c}}=eB/m^{*}$ is the cyclotron frequency in the
magnetic field $B$.
In the Fock--Darwin basis, the Hamiltonian is
\begin{equation*} H_{0}=\sum_{nm}\epsilon_{nm}|nm\rl nm|,
\end{equation*} where
\begin{equation*} \epsilon_{nm}=\hbar\omega_{\mathrm{B}}(2n+|m|+1)
-\frac{1}{2} \hbar\omega_{\mathrm{c}}m.
\end{equation*}

The second term in Eq.~(\ref{ham}) describes a weak anisotropy
(ellipticity) of the confinement potential and has the form
\cite{jacak03a} 
\begin{eqnarray} H_{\mathrm{a}} & = &
\frac{\beta}{2}m^{*}\omega_{0}^{2}(x^{2}-y^{2}) \nonumber  \\ & = &
\frac{\beta}{2} \frac{\hbar \omega_{0}^{2}}{\omega_{\mathrm{B}}}
\sum_{nm,n'm'}V_{(nm)(n'm')}|nm\rl n'm'|\nonumber,
\end{eqnarray} where $\beta$ is a dimensionless parameter and the
non-vanishing matrix elements in the basis of Fock--Darwin states are
\begin{equation*} V_{(0\bar{2})(00)}=\frac{\sqrt{2}}{2},\quad
V_{(0\bar{2})(10)}=\sqrt{2},\quad V_{(0\bar{1})(01)}=1,
\end{equation*} with the symmetries
$V_{(nm)(n'm')}=V_{(n'm')(nm)}=V_{(n\bar{m})(n'\bar{m'})}$.  Here and
throughout the paper, a bar over a number denotes a minus sign.
This anisotropy term leads to the splitting of the $p$-shell states
(in zero magnetic field) given by $\Delta E_p = \beta \hbar \omega_0$
which can be read off the spectral position of the $p$-shell states.

The third part in Eq.~(\ref{ham}) accounts for non-parabolicity of the
confining potential
\begin{equation*} H_{\mathrm{np}}= - \frac{1}{2} \hbar \omega_0 \chi
\left ( \frac{r_{\bot}}{l_0} \right )^4,
\end{equation*} where $ l_{0}$ is the in plane confinement length in
the absence of magnetic field and $\chi\ll1$ is a positive parameter
defining the strength of non-parabolicity.

In the basis of Fock--Darwin states, the non-parabolicity term reads
\begin{equation*} H_{\mathrm{np}} = - \chi \hbar \omega_0
\sum_{(n,m) (n'm')} \mathcal{V}_{(nm)(n'm')}|nm \rangle\! \langle
n'm'|,
\end{equation*} where non-vanishing matrix elements are
\begin{eqnarray}\label{wzor:anharm33} \mathcal{V}_{(00)(00)} & = & 4 ,
\nonumber\\ \mathcal{V}_{(00)(10)}=\mathcal{V}_{(10)(0,0)} & = & 8 ,
\nonumber \\ \mathcal{V}_{(01)(01)}=\mathcal{V}_{(0\bar 1)(0 \bar 1)}
& = & 12 , \nonumber \\ \mathcal{V}_{(02)(02)}=\mathcal{V}_{(0\bar
2)(0\bar 2)} & = & 24 ,\nonumber \\ \mathcal{V}_{(10)(10)} & = &
28. \nonumber
\end{eqnarray}

The electron-phonon coupling Hamiltonian has the form \cite{jacak03a}
\begin{equation*} H_{\mathrm{int}}=\sum_{nmn'm'}|nm\rl n'm'| \sumk
F_{(nm)(n'm')}(\kk)\bk+\mathrm{H.c},
\end{equation*} where
\begin{eqnarray} \lefteqn{F_{(nm)(n'm')}(\kk) =}\nonumber\\ &&
\sqrt{\frac{\hbar\Omega}{2v\varepsilon_{0}\tilde{\varepsilon}}}
\frac{e}{k}f_{(nm)(n'm')}(q) e^{-q^{2}-\xi^{2}}e^{i(m'-m)\phi}\nonumber.
\end{eqnarray} Here we write the wave vector as
$\kk=(k_{\bot}\cos\phi,k_{\bot}\sin\phi,k_{z})$ and introduce the
short-hand notation $q=k_{\bot}l_{\mathrm{B}}/2$, $\xi=k_{z}l_{z}/2$;
$\Omega$ is the frequency of LO phonons at $\kk=0$
($\hbar\Omega=36.7~\mathrm{meV}$), $v$ is the normalization volume for
the phonon modes, $\varepsilon_{0}$ is the vacuum permittivity, and
$\tilde{\varepsilon}= (1/{\varepsilon_\infty} -
1/{\varepsilon_\mathrm{s}})^{-1}=70.3$ is the effective 
dielectric constant (again, the values correspond to GaAs). Note
that the Gaussian cut-off at $k\sim 1/l_{\mathrm{B}}$ restricts the
coupling only to long-wavelength modes, so the frequency of LO phonons
can be replaced by its value at the center of the Brillouin zone
(dispersionless approximation).  The coupling functions have the
general symmetry
\begin{equation*} 
F_{(nm)(n'm')}(\kk)=F^{*}_{(n'm')(nm)}(-\kk),
\end{equation*} 
while the $f$ functions for our choice of basis
states satisfy
\begin{equation}
\label{eq:fnm} 
f_{(nm)(n'm')}(q)=f_{(n'm')(nm)}(q)
=f_{(n'\bar{m}')(n\bar{m})}(q).
\end{equation} 
The functions $f_{(nm)(n'm')}(q)$ are listed in
Tab.~\ref{tab:f}. It may be interesting to note that the functions $F$
are not linearly independent. This follows, e.g., from the linear
dependence of the subset of $f$ functions
$\{f_{(00)(00)},f_{(01)(01)},f_{(10)(00)}\}$, all of which correspond
to $m' - m =0$. The lack of linear independence reduces the number of
collective modes needed to represent the system.

\begin{table}[tb]
\begin{displaymath}
\begin{array}{|l|*{4}{c}|} nm & 00 & 01 & 10 & 02 \\ \hline 00 & 1
&-iq &-q^{2} &-q^{2}/\sqrt{2}\\ 0\bar{1}& -iq&-q^2 & i(q^3-q) &
iq^3/\sqrt{2}\\ 01 & -iq&1-q^2 & i(q^3-q) & i(q^3-2q)/\sqrt{2}\\ 10
&-q^2&i(q^3-q)&1-2q^2+q^4&(q^4-2q^2)/\sqrt{2}\\
0\bar{2}&-q^2/\sqrt{2}&iq^3/\sqrt{2}&(q^4-2q^2)/\sqrt{2}& q^4/2 \\ 02
&-q^2/\sqrt{2}&i(q^3-2q)/\sqrt{2}&(q^4-2q^2)/\sqrt{2}&1-2q^2+q^4/2 \\
\end{array}
\end{displaymath}
\caption{\label{tab:f}Functions $f_{(nm)(n'm')}(q)$ used in the
definition of the coupling constants. Functions not listed here can be
found using Eq.~(\ref{eq:fnm}). }
\end{table}

The last contribution to the Hamiltonian,
\begin{equation*} H_{\mathrm{ph}}=\hbar\Omega\sumk\bkd\bk
\end{equation*} describes free, dispersionless LO phonons.

\section{The numerical approach}
\label{sec:numerics}

In this Section, we describe the general framework for the numerical
diagonalization of the carrier-phonon Hamiltonian. Then, in Section
\ref{sec:numres} we present the results for a few classes of systems.

\begin{table}[tb]
\begin{displaymath}
\begin{array}{|l|*{3}{c}|} & \alpha=A& \alpha=B &\alpha=C\\ \hline M=0
&-q^{2}/\sqrt{x_{4}} &
\frac{x_{4}q^{4}-x_{6}q^{2}}{\sqrt{x_{4}^{2}x_{8}-x_{4}x_{6}^{2}}} &
(1-a_2q^2+a_4q^4)/\sqrt{c}\\ M=\pm1 & iq/\sqrt{x_{2}} &
\frac{i(x_{2}q^{3}-x_{4}q)}{\sqrt{x_{6}x_{2}^{2}-x_{4}^{2}x_{2}}} & \\
M=\pm2 &-q^{2}/\sqrt{x_{4}} &
\frac{x_{4}q^{4}-x_{6}q^{2}}{\sqrt{x_{4}^{2}x_{8}-x_{4}x_{6}^{2}}} &
\\ M=\pm3 & iq^{3}/\sqrt{x_{6}} & & \\ M=\pm4 & q^{4}/\sqrt{x_{8}} & &
\end{array}
\end{displaymath}
\caption{\label{tab:fi}Functions $\phi_{M\alpha}(q)$ used in the
definition of collective modes.}
\end{table}

Our approach to the diagonalization of the Hamiltonian (\ref{ham}) is
based on the collective mode representation of the LO phonons
\cite{stauber00}. We use the basis of the electron subsystem composed
of up to 6 lowest Fock--Darwin states (3 lowest energy shells,
$2n+|m|+1\le 3$).  For this truncated basis, we define 14 collective
phonon modes which are needed to exactly represent the carrier-phonon
coupling in the dispersionless approximation,
\begin{equation} B_{M\alpha}=\sqrt{\frac{l_{\mathrm{B}}}{v}} \sum_k
\frac{1}{k} \phi_{M\alpha}(q) e^{-q^{2}-\xi^{2}+iM \phi}b_{k},
\label{B}
\end{equation} where $\alpha=A,B,C$ labels different modes with the
same angular momentum $M$ and the functions $\varphi_{M\alpha}(q)$ are
listed in Tab.~\ref{tab:fi}. For an axially symmetric dot, the
appropriate functions are expressed in terms of the shape-dependent
parameters (defined for $l$ even)
\begin{equation*} x_{l}=\frac{1}{4\pi^{3}}\int d^{3}q
\frac{q_{\bot}^{l}}{q^{2}}
\exp\left[-2\left(q_{\bot}^{2}+\frac{l_{z}^{2}}{l_{\mathrm{B}}^{2}}q_{z}^{2}
\right)\right].
\end{equation*} We define also
\begin{eqnarray*} a_{2} & = & \frac{x_2 x_4 x_8 - x_4^2 x_6}{{x_4^2
x_8 - x_4 x_6^2}},\\ a_{4} & = & \frac{x_2 x_4 x_6 - x_4^3}{{x_4^2 x_8
- x_4 x_6^2}},\\ c & = & x_0 - 2a_{2}x_{2} + 2a_{4}x_{4}+a_2^2 x_4 - 2
a_2 a_4 x_6 + a_4^2x_6.
\end{eqnarray*} The numbers $x_{l}$ can be found exactly in the limit
of a strong vertical confinement, $l_{z}/l_{\mathrm{B}}\to 0$,
\begin{equation*} x_{l}\to \frac{(l-1)!!}{2^{l+2}\sqrt{2\pi}}.
\end{equation*} These limiting values are collected in
Tab. \ref{tab:xl} and compared with those for
$l_{z}/l_{\mathrm{B}}=0.1$. The leading order correction is
$O(l_{z}^{2}/l_{\mathrm{B}}^{2})$.

\begin{table}[tb]
\begin{displaymath}
\begin{array}{|c|r *{1}{c}|}
&\multicolumn{1}{c}{~~l_z/l_{\mathrm{B}}=0~~}
&~~l_z/l_{\mathrm{B}}=0.1~~\\ \hline x_2 &
\frac{1}{16\sqrt{2\pi}}\approx 0.0249 & 0.0221\\ x_4 &
\frac{3}{64\sqrt{2\pi}}\approx 0.0187 & 0.0159\\ x_6 &
\frac{15}{266\sqrt{2\pi}}\approx 0.0234 & 0.0193\\ x_8 &
\frac{105}{1024\sqrt{2\pi}}\approx 0.0409 & 0.0329\\
\end{array}
\end{displaymath}
\caption{\label{tab:xl}Comparison between numbers $x_l$ calculated for
strong confinement limit $l_z/l_{\mathrm{B}} \to 0$ and for a
realistic value $l_z/l_{\mathrm{B}} = 0.1$ at $B=0$.}
\end{table}

With the definition (\ref{B}), the collective operators
$B_{M\alpha},B^{\dag}_{M\alpha}$ satisfy the usual bosonic commutation
relations,
$[B_{M\alpha},B^{\dag}_{M'\alpha'}]=\delta_{MM'}\delta_{\alpha\alpha'}$
(that is, we follow the standard approach of orthogonalized modes
\cite{stauber00}, although an alternative approach is also possible
\cite{obreschkow07}).  In terms of the collective modes, the
interaction Hamiltonian reads
\begin{eqnarray} H_{\mathrm{int}} & = & \sqrt{\frac{\hbar \Omega
e^2}{2l_B\varepsilon_0 \tilde \varepsilon}}
\sum_{nmn'm'}\sum_{\alpha}|nm\rl n'm'|  \nonumber\\ && \times
\gamma_{(nm)(n'm')\alpha} B_{m'-m,\alpha}  +\mathrm{H.c.}\nonumber,
\end{eqnarray} where the coupling constants
$\gamma_{(nm)(n'm')\alpha}$ are collected in Tab.~\ref{tab:gamma}.
The mode $B_{0\mathrm{C}}$ couples only to the unit operator on the
restricted electron subspace, $\mathbb{I}=\sum_{mn}|mn\rl mn|$, and
can be discarded from the discussion \cite{stauber00}.

\begin{table*}[tb]
\scriptsize
\begin{displaymath}
\begin{array}{|r|*{4}{c}|} nm~\alpha & 00& 01 & 10 & 02 \\ \hline 00~A
& (a_{4}x_{6}-a_{2}x_{4})/\sqrt{x_{4}} & -\sqrt{x_{2}} & \sqrt{x_{4}}
& \sqrt{x_{4}/2} \\ B &{-a_{4}\sqrt{x_{8}- {x_{6}^{2}}/{x_{4}}}} & & &
\\ C & \sqrt{c} & & & \\ 0\bar{1}~A & -\sqrt{x_2} & \sqrt{x_4} &
({x_4}/{x_2} -1) \sqrt{x_2} & \sqrt{x_6 / 2}\\ B & & &{\sqrt{x_6 -
{x_4^2}/{x_2}}} & \\ 01~A & - \sqrt{x_2} &a_4 x_6/\sqrt{x_4} - (a_2
+1)\sqrt{x_4} & {{{x_4}}/{\sqrt{x_2}} - \sqrt{x_2}} &
{{x_4}/{\sqrt{2x_2}} - \sqrt{2x_2}}\\ B &
&-a_{4}\sqrt{x_{8}-{x_{6}^{2}}/{x_{4}}} &{\sqrt{x_6 - {x_4^2}/{x_2}}}
&{\sqrt{{x_6}/{2} - {x_4^2}/{2x_2}}}\\ C & & \sqrt{c} & & \\ 10~A
&\sqrt{x_4} & {{x_4}/{\sqrt{x_2}} - \sqrt{x_2}}
&(2-a_2)\sqrt{x_4}+(a_4-1)x_6/\sqrt{x_4} &{\sqrt{2x_4}-
{x_6}/{\sqrt{2x_4}}}\\

B & &{\sqrt{x_6 - {x_4^2}/{x_2}}} & {(1-a_4)}/{x_4}\sqrt{x_{8}-
{x_{6}^{2}}/{x_{4}}} &{\sqrt{{x_8}/{2} - {x_6^2}/{2x_4}}}\\ C & & &
\sqrt{c} & \\

0\bar{2}~A& \sqrt{x_4/2} &{\sqrt{x_6/2}} & {\sqrt{2x_4}-
{x_6}/{\sqrt{2x_4}}} & \sqrt{x_8}/2\\ B & & & {\sqrt{{x_8}/{2} -
{x_6^2}/{2x_4}}} &\\

02~A & \sqrt{x_4/2} & {{x_4}/{\sqrt{2x_2}} - \sqrt{2x_2} } &
\sqrt{2x_4}- {x_6}/{\sqrt{2x_4}} &
(2-a_2)\sqrt{x_4}+(a_4-\frac{1}{2})x_6/\sqrt{x_4} \\

B & & {\sqrt{{x_6}/{2} - {x_4^2}/{2x_2}}} &
{\sqrt{{x_8}/{2}-{x_6^2}/{2x_4}}} &
\sqrt{x_{8}-x_{6}^{2}/x_{4}}(\frac{1}{2}-a_4)/x_4 \\ C & & &
&\sqrt{c}\\

\end{array}
\end{displaymath}
\normalsize
\caption{\label{tab:gamma}Coupling constants
$\gamma_{(nm)(n'm')\alpha}$ for the collective LO modes. Definitions
as in Tab.~\ref{tab:fi}. The values not listed in the table can be
reproduced form the relation
$\gamma_{(n\bar{m})(n'\bar{m}')\alpha}=\gamma_{(nm)(n'm')\alpha}.$}
\end{table*}

The Hamiltonian (\ref{ham}) is then diagonalized numerically,
including states with up to 3 phonons, which yields a computational
basis of 4080 states. The relevance of 4-phonon states is discussed in
the Appendix.

\section{Numerical Results}
\label{sec:numres} 
In this section, we
present the results of a numerical investigation on the second-order
resonant polarons. First, we study the system in the absence of a
magnetic field. We calculate the system spectrum as a function of the
separation between the unperturbed electron energy levels (that is,
indirectly, on
the QD size). We model also the intraband absorption spectrum of an
inhomogeneously broadened ensemble of QDs, where the polaron resonance
is clearly manifested. Next, we calculate the spectrum of a single QD
with a fixed size as a function of an external axial magnetic field and
compare the results to the existing experimental data.

When plotting the obtained polaron spectra, we present only the states with
a sufficient transition probability, that way spurious uncoupled states 
(for the truncation at the 3-phonon level) are not included. Since we focus
on the second-order resonance, all important couplings are contained in the
model.

\subsection{Size-dependent polaron spectrum} 

We first consider the system
without a magnetic field and focus on the resonance between the
unperturbed excited electronic states $|0\pm1\rangle$ and a set of
two-phonon states $|00,2\mathrm{ph}\rangle$. Here and throughout the
paper notation of this form stands for all two-phonon states with the
electronic part in the ground state: $|00\rangle_{\mathrm{el}} \otimes
(B_{M\alpha}^\dag B_{M'\alpha'}^\dag |0\rangle_{\mathrm{ph}})$.  Any
other group of states with different number of phonons is denoted in
a similar way. The quantities of interest in the present
calculation are energies of the system eigenstates and the photon
absorption probability for an intraband transition between the ground
state and a given excited state. The former are obtained directly for
the numerical diagonalization, while the latter, for an excited state
$|\kappa\rangle$, is calculated according to
\begin{equation}
\label{eq:transition} 
|\mu_\kappa|^2 = \left |\langle
\kappa | \hat d^{(-)}_{\lambda} |\Psi_{\mathrm{G}} \rangle \right |^2,
\end{equation} 

where $\Psi_G$ denotes the numerically calculated
ground state and  $\hat{d}^{(-)}_\lambda$ is the negative frequency
part of the dipole moment operator, which depends on the polarization
$\lambda$ of the optical wave 
exciting the system.  For the $\sigma_{\pm}$ circular polarizations
one has (upon truncation to our computational space)
\begin{eqnarray*}
\hat d^{(-)}_{+} & \propto & (|01\rl 00|
+ \sqrt{2} |02 \rl 01| + |10 \rl 0 \bar{1}| + |01\rl 10| \nonumber \\
&& + \sqrt{2} |0 \bar 1 \rl 0 \bar 2 | + |00 \rl 0 \bar 1|)
\otimes\mathbb{I}_\mathrm{ph}
\label{eq:dipoleplus} 
\end{eqnarray*}
and
\begin{eqnarray*}
\hat d^{(-)}_{-} & \propto & (|0\bar{1}\rl 00|
+ \sqrt{2} |0\bar{2} \rl 0\bar{1}| + |10 \rl 0 1| + |0\bar{1}\rl 10|\nonumber \\
&& + \sqrt{2} |0 1 \rl 0 2 | + |00 \rl 01|)
\otimes\mathbb{I}_\mathrm{ph}, 
\label{eq:dipoleminus} 
\end{eqnarray*}
where $\mathbb{I}_\mathrm{ph}$ is the identity operator
on the phonon subsystem.
For the linear polarization along the $x$ and $y$ axes, the dipole
moment operator is
\begin{eqnarray*}
\label{eq:dipolemom} 
\hat d^{(-)}_x \propto d^{(-)}_{+}+d^{(-)}_{-},\quad
\hat d^{(-)}_y \propto d^{(-)}_{+}-d^{(-)}_{-}.
\end{eqnarray*} 

\begin{figure}[tb]
\includegraphics[width=85mm]{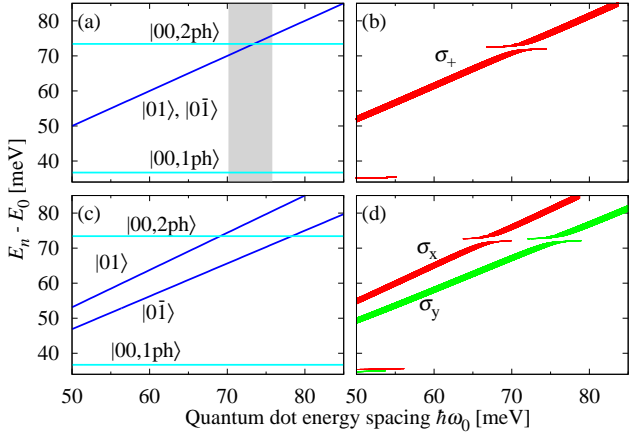}
\caption{\label{fig:heff2}(a,c) Dependence of the unperturbed
eigenenergies (without electron-phonon coupling) on characteristic
energy of the QD. (b,d) The QD energy spectrum with electron-phonon
interaction included. Panels (a) and (b) refer to a cylindrically
symmetric dot, while (c) and (d) present the spectra for an elliptical
dot. In (b) and (d) the line thickness is proportional to the
intraband transition probability $|\mu_{k}|^{2}$, the lines are labeled by the
polarization of the exciting light. Only states for which
$|\mu_{k}|^{2}>0.1$ are plotted.}
\end{figure}

We investigate two cases: an isotropic QD and an anisotropic one
(Fig.~\ref{fig:heff2}).  In Figs.~\ref{fig:heff2}(a) and
\ref{fig:heff2}(c), we have presented states from the $p$ shell without
phonons as well as one-phonon and two-phonon states with the electronic
part in the ground state ($|00,1\mathrm{ph}\rangle$,
$|00,2\mathrm{ph}\rangle$). The energies are shown relative to the
system ground state. All the other unperturbed states lie in a
higher energy range. In the case of an isotropic dot, in the absence
of the electron-phonon coupling, for a certain value of the electron
level spacing (shaded area in Fig.~\ref{fig:heff2}a), $p$-shell states
($|0\pm 1\rangle$) intersect with the group of two-phonon states with
the electron in its ground state ($|00,2\mathrm{ph}\rangle$).

When the electron-phonon coupling is included, this intersection turns
into an avoided crossing pattern, as shown for an isotropic dot in
Fig.~\ref{fig:heff2}(b). Importantly, there are no direct matrix
elements coupling those states and the coupling between the $p$-shell
zero-phonon states and the two-phonon states is mediated through
one-phonon states. 
Even though this coupling is of the second order the anticrossing is
quite strong and its width is 1.75 meV. 

If anisotropy is included, the degeneracy of states $|01\rangle$ and
$|0\bar1\rangle$ is lifted and two lines in the polaron spectrum are
observed for different linear polarizations of the incident light. As
can be seen on Figs.~\ref{fig:heff2}(c) and 
\ref{fig:heff2}(d), there are two intersections
[Fig.~\ref{fig:heff2}(c)] resulting in two anticrossings
[Fig.~\ref{fig:heff2}(d)] which become visible in the absorption
spectrum depending on the polarization. 

In the low energy range [36 meV, Figs. \ref{fig:heff2}(b),
\ref{fig:heff2}(d)] states with small but noticeable transition
probability can be observed. Their existence is due to the first order
coupling leading to some transfer of the oscillator strength from the
$p$-shell states to 1-phonon states
$|00,1\mathrm{ph}\rangle$.

The position of the resonance in Figs.~\ref{fig:heff2}(b) and
\ref{fig:heff2}(d) is shifted with respect to the intersection of
decoupled states [Figs.~\ref{fig:heff2}(a) and \ref{fig:heff2}(c)].
The center of the resonance for the interacting case is located at a lower QD
energy spacing (71 meV) than the intersection point between
non-interacting states $|00,2\mathrm{ph}\rangle$ and $|0\pm1\rangle$
(73.4 meV). Such a behaviour results from the presence of other states
directly coupled to zero- and two- phonon lines. The most important
states influencing the position of the resonance are 3-phonon states
$|00,3\mathrm{ph}\rangle$ which are relatively close to the
$|00,2\mathrm{ph}\rangle$ states and effectively reduce their energy.

Since the second-order resonance is relatively strong it should also
be visible in the absorption spectra of inhomogeneously broadened
ensembles. This is discussed in Sec. \ref{sec:ensemble}. More insight to
the structure of the second-order resonant polarons can be obtained
using the effective Hamiltonian approach which is presented in
Sec.~\ref{sec:resHeff}.

\subsection{Polaron resonance in the ensemble absorption} 
\label{sec:ensemble}
In this
section, intraband absorption spectra of an inhomogeneously broadened
QD ensemble are calculated. QD sizes in self--assembled QD ensembles
are always given by some distribution. We take this inhomogeneity of
sizes into account and theoretically investigate the 
ensemble intraband absorption
spectrum in the area of the two phonon resonance.  

The QDs are
parametrized by their energy spacing $\hbar \omega_0^{(i)}$, where $i$
labels dots in the ensemble.
The distribution of QD energies is assumed to be described by a
Gaussian function 
\begin{equation*}
\label{eq:gauss} 
f_{\overline{\epsilon},\sigma}(\hbar \omega_0^{(i)})=
\frac{1}{\sigma \sqrt{2\pi}} e^{-\frac{(\hbar \omega_0^{(i)}-\overline{\epsilon})^2}{2
\sigma^2}}, 
\end{equation*} 
where $\overline{\epsilon}=\hbar \overline{\omega_0}$
and $\sigma$ are the mean transition energy and its standard
deviation, respectively.  Those parameters will be chosen in such a way
that the second order resonance is located in the high-energy tail of
the QD size distribution.

In order to construct high quality absorption spectra, we calculate
the polaronic states for up to $N=10^5$ QDs with the transition energies
$\hbar \omega_0^{(i)}$ uniformly distributed over a sufficiently broad
range. From these numerical results, the absorption spectrum for the
requested polarization is calculated according to
\begin{equation*}
\label{eq:intqd} 
I(\hbar \omega) \propto \sum_{i=1}^N
\sum_{\kappa=1}^{N_{\mathrm{ev}}} \delta (\hbar \omega -
\mathcal{E}_{i\kappa}) |\mu_\kappa|^2 f_{\overline{\epsilon},
\sigma^2}(\hbar \omega_0^{(i)}) ,
\end{equation*} 
where $\hbar \omega$ is the energy of the absorbed
photon, $N$ and $N_{\mathrm{ev}}$ are the number of QDs used in
calculations and the number of eigenvalues for $i$--th QD,
respectively, ${\mathcal{E}_{i\kappa}}$ stands for $\kappa$--th
eigenenergy of $i$--th quantum dot and $|\mu_\kappa|^{2}$ is
given by Eq.~(\ref{eq:transition}).

\begin{figure}[tb]
\includegraphics[width=85mm]{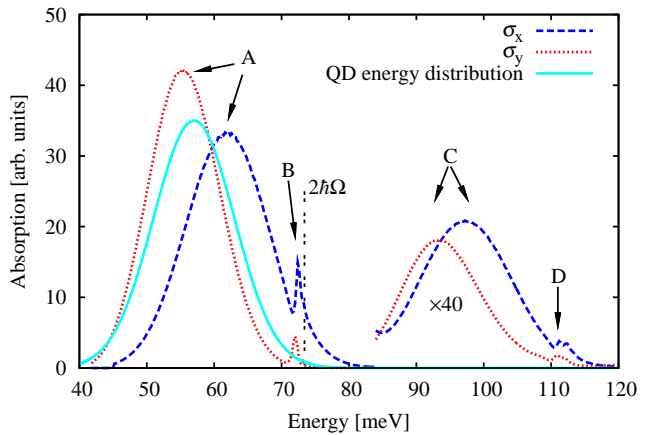}
\caption{(Color online) Absorption spectra for two different
linear polarizations ($\sigma_X$, $\sigma_Y$) with anisotropy
and without non-parabolic corrections. Distribution of characteristic
energies $ \hbar \omega_0^{(i)}$ is also presented. The mean energy
and its standard deviation are taken respectively as $\hbar
\overline{\omega_0} = 57~\mathrm{meV}$ and $\sigma =
6~\mathrm{meV}$. Labels denote features discussed in the text. The
dotted vertical line marks the 2-LO phonon energy. The right part of the
plot is scaled up, as marked.}
\label{fig:test4}
\end{figure} 

The absorption spectrum in the case of a polaron in an anisotropic,
parabolic confinement potential is presented in
Fig.~\ref{fig:test4}. The polaronic feature is clearly manifested for
energies close to the energy of two LO phonons (feature B in the
plot). Additionally, a phonon replica (C) of the main absorption peak
is visible for higher energies,
including the resonance feature (D). The 
latter is slightly broadened and consists of two peaks (D). It is
worth mentioning that the main absorption feature (A), the resonant
feature (B), as well as the phonon-replica (C) are reproduced
correctly in a diagonalization with up to 3-phonon states included. On
the other hand, the replica of the resonant feature (D) is modelled
correctly only if 4-phonon states are taken into account (see Appendix
A).
The absorption features for the $\sigma_X$ and $\sigma_Y$
polarizations are not related by symmetry with respect to the average QD energy
($\overline{\epsilon}$). This effect is due to coupling to a lower lying
group of one-phonon states, which moves the relevant eigenvalues to
higher energies.

\begin{figure}[tb]
\begin{center}
\includegraphics[width=85mm]{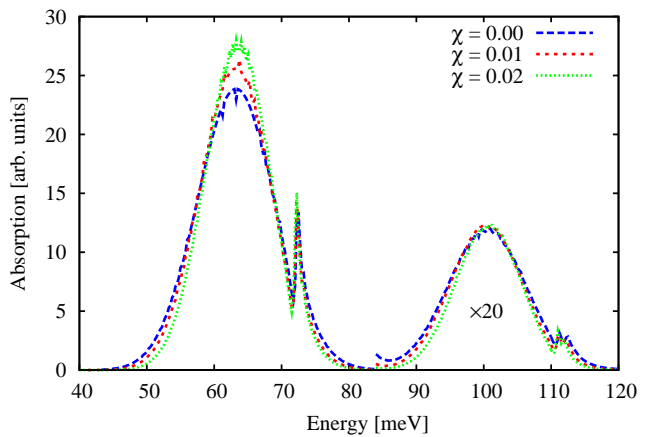}
\caption{(Color online) Absorption spectra on ensembles of isotropic
QDs for different non-parabolic corrections. Standard deviation $\sigma
= 6~\mathrm{meV}$, the mean energy $\hbar\overline{\omega_0}= 62$,
$67.1$ and $72.8~\mathrm{meV}$ respectively for ascending values of
$\chi$.}
\label{fig:abs0Banh}
\end{center}
\end{figure}

The position of the second-order polaronic feature is slightly lower
than the energy of two LO phonons $2\hbar \Omega$. This is
mostly due to the interaction with a group of 3-phonon states
$|00,3\mathrm{ph}\rangle$ which effectively reduces the energy of the
2-phonon line. Also the position of the polaronic feature varies
slightly with the average value $\hbar \overline{\omega_0}$ of the QD
energy distribution. For the average energies higher than the energy
of two LO phonons the resonance position may shift towards lower
energies. However such a shift is rather small and does not exceed
$1~\mathrm{meV}$. The position of the second-order resonant feature 
in the ensemble absorption is
thus not fixed and may vary with distribution of QDs in an ensemble.

Although the potential confining electrons in a QD is often considered
parabolic, more realistic modelling must take into account its
non-parabolicity.  For the sake of simplicity, calculations for a
non-parabolic confining potential are performed for isotropic QDs, that
is, the case where the anisotropic term $H_{\mathrm{a}}$ in
Eq.~(\ref{ham}) is discarded.  In Fig.~\ref{fig:abs0Banh}, we present
absorption spectra for different strengths of non-parabolicity
$\chi$. Distribution of the QD characteristic energies is tuned so that
the main absorption peak has the same position.  
Since in a non-parabolic dot higher levels are closer to the resonant
group of states we expected that the resonance may be shifted down by
coupling to these states. However, nothing like this is observed: 
Even for a strong
non-parabolicity, only a negligible change in the position of the
second-order resonance is observed.

\subsection{Magnetopolaron resonances} 

\begin{figure}[tb]
\includegraphics[width=85mm]{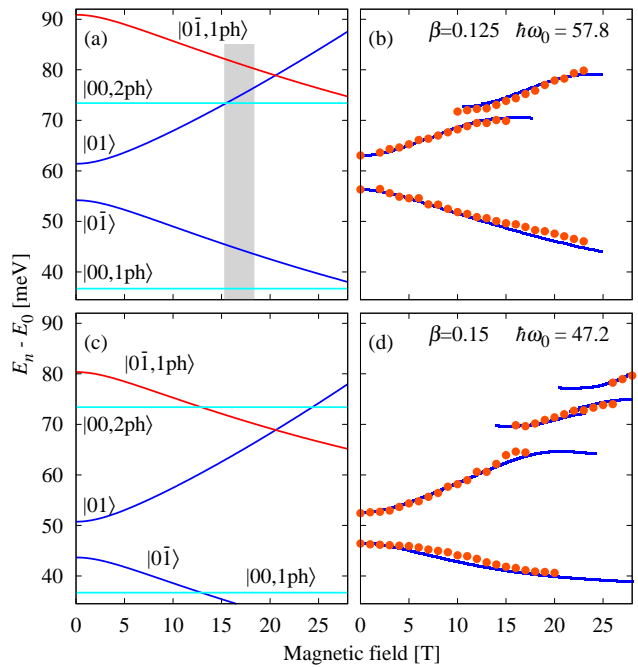}
\caption{\label{fig:sample3}(Color online) (a,c) Dependence of the
unperturbed eigenenergies (without electron-phonon coupling) on
magnetic field. (b,d) The QD energy spectrum in magnetic field with
electron-phonon interaction included. Results obtained with numerical
diagonalization of full Hamiltonian (lines) compared with experimental
results taken from Refs.~\onlinecite{hameau99,hameau02} (points).}
\end{figure}

In this section, we investigate
a single QD in a magnetic field. We consider a magnetopolaron
resonance, that is, the case of bringing the state $|01\rangle$ to
resonance with 2-phonon states $|00,2\mathrm{ph}\rangle$ using energy
level shifts in an
external magnetic field.  In Figs.~\ref{fig:sample3}(a) and
\ref{fig:sample3}(c), we present the spectrum of a single isotropic QD
without electron-phonon coupling in a perpendicular magnetic field,
for two different characteristic energies $\hbar \omega_0$. In both
Figs. \ref{fig:sample3}(a) and \ref{fig:sample3}(c) there are several
intersections. The first one, between the state $|0\bar1\rangle$ and a
group of 1-phonon states, corresponds to the first order
resonance. States intersecting in the upper part of the charts are the
purely electronic excited state $|01\rangle$, 2-phonon states with the
electronic part in the ground state $|00,2\mathrm{ph}\rangle$, and
1-phonon states with an excited electronic part
$|0\bar1,1\mathrm{ph}\rangle$. Depending on the size of the QD, those
intersections may appear at different magnitudes of the magnetic field
and in different order.

For the present discussion, the intersection between the purely
electronic state $|01\rangle$ and a group of 2-phonon states
$|00,2\mathrm{ph}\rangle$ is of interest [shaded area in
Fig.~\ref{fig:sample3}(a)]. In a strong magnetic field, one group of
states mediating the interaction ($|0\bar1,1\mathrm{ph}\rangle$)
between 0- and 2-phonon states is much closer to the resonance than at
$B=0$. States $|0\bar1,1\mathrm{ph}\rangle$ are directly coupled to
both relevant groups of states. Since the direct interaction is much
stronger than indirect one, the resonance between zero-phonon and
two-phonon states can be strongly intermixed with 1-phonon
states. This is especially the case for QD sizes in between those in
Figs.~\ref{fig:sample3}(a) and \ref{fig:sample3}(c).

We compare the results of numerical diagonalization with experimental
ones \cite{hameau99,hameau02} obtained for 2 samples differing in QDs
sizes. The diagrams presented in Figs.~\ref{fig:sample3}(b) and
\ref{fig:sample3}(d) consist of several lines that represent
transitions to different excited states. The splitting of the $p$-shell 
states allows one to uniquely determine the anisotropy parameter
 $\beta$, while the confinement energy $\hbar \omega_0$ can be directly 
read off the spectral position of these $p$-shell states. 
For low magnetic fields, two
lines with an opposite Zeeman shift emerge. At high magnetic fields, the
lower lying line strongly interacts with 1-phonon states
$|00,1\mathrm{ph}\rangle$, that is, a strong shift towards higher
energies is observed. For smaller QDs (and thus higher energy
spacing $\hbar \omega_0$), for a magnetic field in the range 10-15~T,
a clean second-order resonance between $|01 \rangle$ and
$|00,2\mathrm{ph}\rangle$ states can be observed
[Fig.~\ref{fig:sample3}(b)]. For a bigger QD
[Fig.~\ref{fig:sample3}(d)], when the upper line comes close to 1- and
2-phonon states, two resonances appear. One may assume that the first
anticrossing in the upper part of this chart is due to the direct
interaction between the states $|0\bar1,1\mathrm{ph}\rangle$ and
$|01\rangle$ whereas the second one is a second-order resonance
between the states $|00,2\mathrm{ph}\rangle$ and $|01\rangle$. However
it must be noted that all these states are relatively close to each
other and both, first and second order polarons can be strongly
intermixed.

Previous studies \cite{hameau99,hameau02,jacak02a,jacak03a,jacak02f}
suggested 
that the electron-phonon interaction, measured by the dimensionless
Fr\"ohlich constant 
\begin{equation*}
\alpha_F = \frac{e^2}{\epsilon_0 \tilde \epsilon}
\sqrt{\frac{m^{*}}{2\hbar^3\Omega}}, 
\end{equation*}
needs to be tuned in order to
conform with experiment. As a result, the Fr\"ohlich constant, which
in principle depends only on material properties, was used as an
adjustable 
parameter.  Our present results show that such a treatment is not
necessary.  As can be seen in
Figs. \ref{fig:sample3}(b) and \ref{fig:sample3}(d), both resonance
positions and widths are reproduced without
the need to enhance the Fr\"ohlich constant after a sufficient number of
electron shells and phonon modes has been included in the computational
basis. For instance, our numerical solution yields the resonance width
of $3.0$~meV in the case shown in Fig.~\ref{fig:sample3}(b), which
compares quite well with the experimental value of $2.9$~meV. By
performing the diagonalization in restricted bases, we have found out
that the 3-shell, 3-phonon model is actually minimal in the sense that
a further reduction of the basis set leads to incorrect results. In
particular, leaving out the $d$-shell states yields a strongly
underestimated width of the resonance. On the other hand, not
accounting for 3-phonon states moves the resonance to higher magnetic
fields and produces nonexistent polaron branches in the resonance area
(See Ref.~\onlinecite{kaczmarkiewicz08} for more details).

\begin{table}[tb]
\begin{displaymath}
\begin{array}{|c|c|c|} \mathrm{Computational} &
\mathrm{Resonance}&\mathrm{Resonance}\\ \mathrm{basis} &
\mathrm{width~[meV]}&\mathrm{position~[T]}\\ \hline
\mathrm{Full~model} & 3.00 & 13.5 \\ \mathrm{3~phonons,2~shells } &
2.10 & 12.7 \\ \mathrm{2~phonons,3~shells } & 2.98 & 15.2\\
\mathrm{2~phonons,2~shells } & 2.80 & 14.9\\ \hline
\mathrm{Experimental~results} & 2.9 & 12-14 \\
\end{array}
\end{displaymath}
\caption{\label{tab:restrictingbase}Properties of the resonance area
for different truncations of the computational basis, compared to
experimental results \cite{hameau99}.}
\end{table} 

We have also checked to what extent the numerical results depend on
our choice of the $l_{z}/l_{0}$ ratio.  For $l_z/l_0=0.2$, that is,
twice larger 
than used in this paper, only small shift in eigenenergies is
observed (approximately $0.1~\mathrm{meV}$) and the resonance width
decreases to $2.8~\mathrm{meV}$, still very close
to the experimental result of $2.9~\mathrm{meV}$.

\section{Effective Hamiltonian approach}
\label{sec:Heff} 

In the effective Hamiltonian approach, one considers the case of an
unperturbed Hamiltonian $H_0$ having eigenvalues $E_{i\alpha}$ grouped
into well separated manifolds \cite{cohen98}. The states
can be written in the form $|i \alpha \rangle$, where the Latin
indices denote different states within a manifold while Greek indices
refer to different manifolds.  The corresponding energies of the
unperturbed system are $E_{i\alpha}$.  Grouping into a manifold means
that
\begin{equation*}
\label{wzor:Heffll} 
| E_{i\alpha} - E_{j\alpha}| \ll
|E_{i\alpha} - E_{j\beta} |~~ \mathrm{for}~ \alpha \neq
\beta,
\end{equation*} 
i.e., energy separation between states from different
manifolds, is much larger then energy separations within
manifold. Moreover, perturbation-induced coupling between
states from different 
manifolds should be much smaller than the energy separation between these
states,
\begin{equation}\label{wzor:Heffrel} 
|\langle i,\alpha | V |
j,\beta \rangle | \ll |E_{i\alpha} - E_{j \beta}|~~ \mathrm{for}~
\alpha \neq \beta,
\end{equation}
where $V$ is the perturbation.

The eigenvalue problem is treated by a quasi-degenerate perturbation
theory.  The original Hamiltonian $H=H_0 + V$ is transformed
into a new one, $H_{\mathrm{eff}}$, which has no matrix elements
between different groups 
of states (up to the required order of approximation), by means of a
unitary transformation $T=e^{iS}$. The matrix elements of the
operator $S$ and of $H_{\mathrm{eff}}$ can be found iteratively.

The expression for the second-order effective Hamiltonian has the form
\cite{cohen98}
\begin{eqnarray}\label{wzor:Heff} \langle i | H_{\mathrm{eff}}^\alpha
| j \rangle &=& E_{i\alpha} \delta_{ij} + \langle i,\alpha | V
| j,\alpha \rangle +\nonumber \\ && + \frac{1}{2} \sum_{k, \gamma \neq \alpha}
\langle i,\alpha | V | k, \gamma \rangle \langle k, \gamma |
V | j,\alpha \rangle \nonumber \\ && \times \left [
\frac{1}{E_{i \alpha} - E_{k \gamma}} + \frac{1}{E_{j \alpha} - E_{k
\gamma}}\right ] .
\end{eqnarray} 
The first term in Eq.~(\ref{wzor:Heff}) represents the
unperturbed energies, the second one accounts for direct couplings
within a single manifold, and the last one describes the influence of
intermediate states from the different manifolds on the effective
Hamiltonian matrix. This last term represents indirect second-order
couplings between the states of the manifold of interest which result
from the couplings to other manifolds eliminated by the unitary 
transformation $T$.

The effective Hamiltonian method is a powerful tool for calculations
and interpretation of various systems. In the present case of a two-phonon
polaron resonance, it is very helpful since in the resonance area a
purely electronic state and a group of two-phonon states form a well
separated manifold.
The method can be used for a description of both magnetopolaron
resonances and a size-dependent polaron spectrum, though it is more
accurate in the latter case, since the condition for the appropriate relations
between energy spacings and coupling strengths
[Eq.~(\ref{wzor:Heffrel})] is fulfilled in this case
with a greater precision.

If one considers the case of the size dependent spectrum (without a
magnetic field) the energy difference
between states from different manifolds at the point of the resonance
is at least $\hbar \Omega = 36.7~\mathrm{meV}$. On the other hand, if
the purely electronic state $|01\rangle$ is brought to resonance with the
2-phonon line $|00,2\mathrm{ph}\rangle$ using a magnetic field, the energy
separation is significantly lower. For the cases presented in
Figs.~\ref{fig:sample3}(a), \ref{fig:sample3}(c), it is approximately
15 and 30\% smaller, respectively (assuming that the coupled state
$|0\bar{1},1\mathrm{ph}\rangle $ is considered as a member of the
manifold). If we take into consideration that the average direct coupling
between states from different manifolds is about 3~meV, even 30\%
reduction in energy separation might impair the applicability condition.

The effective Hamiltonian approach automatically includes the nested 
coupling polaron structure \cite{obreschkow07}, since from the whole
spectrum of the states it filters out the ones that are coupled to each 
other through intermediate states (for a given truncation of the basis).

\section{Results: effective Hamiltonian}
\label{sec:resHeff} 

In the
following section, we investigate the second-order resonance using the
effective Hamiltonian approach. We consider the case of size dependent
spectra as well as a magnetopolaron resonance.

We apply an appropriate treatment to obtain the effective Hamiltonian
in both cases, although the latter one might be less applicable in the
case of larger QDs (Zeeman tuning brings one-phonon states close to
two-phonon line intermixing first and second order resonances and thus
breaks the condition for sufficient manifolds separation).  Since
the energy separation between different states depends on the size of
the QD each case should be studied individually.

The effective Hamiltonian approach allows us to gain some information
about the second-order polaron structure.  Since we choose relevant
indirectly coupled states only we can get much insight into the mediated
interaction between 0- and 2- phonon states. As the effective Hamiltonian
matrix is much smaller than the full Hamiltonian matrix its diagonalization is much
faster and interpretation of the spectra is easier. Contributions
from different intermediate states can easily be separated and
studied.  In particular, we investigate the influence of $d$-shell and
3-phonon states on the polaron spectra and on the coupling strengths
appearing in the effective Hamiltonian.  We show
that the quasi-degenerate perturbation theory not only allows one to
describe the resonance area in detail, but also explains why both
$d$-shell and 3-phonon states have to be used in order to correctly
model second-order polarons.

\subsection{Polaron resonance at $B=0$} 

In this section, we consider
the effective Hamiltonian approach to the second-order resonance
between states $|01\rangle$ and $|00,2\mathrm{ph}\rangle$ in the
absence of a magnetic field. For the sake of simplicity, we assume that
the QD is isotropic (it
is always possible to introduce anisotropy perturbatively).

Although, for our truncated basis, there are $105$ two-phonon states
only $6$ of them couple indirectly (in the second order approximation)
to the purely electronic state $|01\rangle$.  There is a small number
($12$) of intermediate states which produce nonzero couplings in the
effective Hamiltonian. Their contribution is presented
(grouped by shell and number of phonons) in Tab.~\ref{tab:heffcmp}.
If the numerical values for different shells or different phonon numbers
sum up to zero the relevant states are decoupled. 

\begin{table}[tb]
\begin{displaymath}
\begin{array}{|r|ccc|cc|} \hline
\multicolumn{1}{|c|}{\mathrm{matrix~elements}} &
\multicolumn{5}{|c|}{\mathrm{intermediate~states}}\\ \cline{2-6} & s &
p & d &1\mathrm{ph}&3\mathrm{ph}\\ \hline \langle 01| H_{\mathrm{eff}}
|00,A^2 A^{\bar 1}\rangle & 0 &82.8\%& 17.2\%&1 &0\\ \langle 01|
H_{\mathrm{eff}} |00,A^0 A^1\rangle & \multicolumn{2}{c}{88.9\%} &
11.1\%&1 &0\\ \langle 01| H_{\mathrm{eff}} |00,A^0 B^1\rangle & 0 & 0
& 100\% &1 &0\\ \langle 01| H_{\mathrm{eff}} |00,A^3 A^{\bar 2}\rangle
& 0 & 0 & 100\% &1 &0\\ \langle 01| H_{\mathrm{eff}} |00,A^2 B^{\bar
1}\rangle & 0 & 0 & 100\% &1 &0\\ \langle 01| H_{\mathrm{eff}} |00,B^0
A^1\rangle & 50\%&-50\%& 0 &1 &0\\ \langle 00,A^0 A^1|H_{\mathrm{eff}}
|00,B^0 A^1\rangle & 100\% & 0 & 0 &0.5&-0.5\\ \hline
\end{array}
\end{displaymath}
\caption{\label{tab:heffcmp}Relative influence of different electronic
shells and $n$-phonon states on the effective Hamiltonian couplings in
the limit $l_z/l_B \rightarrow 0$. In the second line, the
contributions from the two shells are taken together as they partly
cancel each other.}
\end{table}

\begin{table}[tb]
\begin{displaymath}
\begin{array}{|c|r|} \hline H_{\mathrm{eff}}~\mathrm{matrix~elements}
& \multicolumn{1}{c|}{\mathrm{coupling~strength}} \\ \hline
\langle01|H_{\mathrm{eff}}|00,A^2A^{\bar1}\rangle &
~~29\sqrt{3}/(768\sqrt{2\pi}) \approx ~~0.0261 \\
\langle01|H_{\mathrm{eff}}|00,A^0A^1\rangle &
~~13\sqrt{3}/(384\sqrt{2\pi}) \approx ~~0.0234 \\
\langle01|H_{\mathrm{eff}}|00,A^0B^1\rangle & -1/(128\sqrt{\pi})
\approx -0.0044 \\ \langle01|H_{\mathrm{eff}}|00,A^3A^{\bar2}\rangle &
-5\sqrt{5}/(256\sqrt{2\pi}) \approx -0.0035\\
\langle01|H_{\mathrm{eff}}|00,A^2B^{\bar1}\rangle & -1/(256\sqrt{\pi})
\approx -0.0022 \\ \hline
\end{array}
\end{displaymath}
\caption{\label{tab:heffcpl}Indirect coupling strengths in the
effective Hamiltonian approach. Coupling strength expressed in units
of ${e^2}/(2l\varepsilon_0\tilde{\varepsilon}) = 32.7~\mathrm{meV}$.}
\end{table} 

As we can see in Tab.~\ref{tab:heffcmp}, taking the $d$-shell into
account not only 
introduces three additional couplings (rows 3 to 5), but also changes
the strength of existing ones (increase 
of over 17\% and 11\% in rows 1 and 2,
respectively).  What is more important, taking $3$-phonon states into account
completely decouples the 2-phonon state $|00,B^0A^1\rangle$ as shown
in row 7. 
Here, $A^{1}$ denotes the phonon mode created by the collective
operator $B^{\dag}_{1A}$ [Eq.~(\ref{B})], etc. 
Although Tab.~\ref{tab:heffcmp} is constructed for the limiting case
of an infinitely flat QD ($l_{z}\to 0$) and at exact resonance ($\hbar
\omega_0 = 2 \hbar 
\Omega$) this decoupling is preserved also for non-flat QDs
and is an effect of equal spacing of the electronic eigenenergies in a
parabolic confining potential.

\begin{figure}[tb]
\includegraphics[width=85mm]{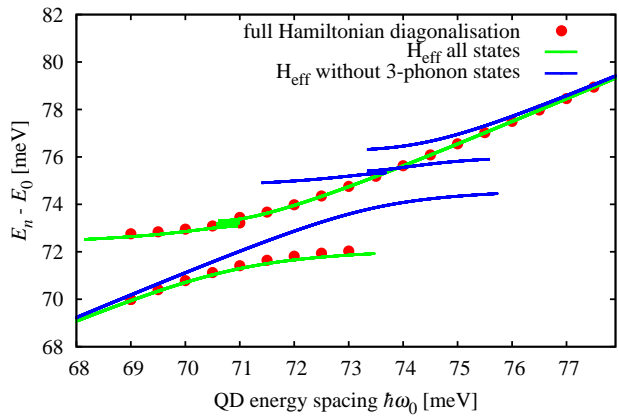}
\caption{\label{fig:porown-heff}(Color online) Comparison between
effective 
Hamiltonian approach and full Hamiltonian diagonalization. Influence
of disregarding 3-phonon states on resonance area also included.}
\end{figure}

The properties of indirect couplings via 1-phonon and 3-phonon states,
revealed by the effective Hamiltonian structure, allow one to
qualitatively understand why a 3-shell, 3-phonon model is required for
correct modelling of the resonant polaron spectrum.
Leaving out 3-phonon states leads to a reconstruction of the spectrum
(additional lines in the theoretically modelled absorption spectrum are
observed, see Fig.~\ref{fig:porown-heff}) since the coupling between
states $|00,B^0A^1\rangle$ and $|00,A^0A^1\rangle$ in the absence of
3-phonon states is relatively strong. For the case of a flat QD, the
coupling strength factor is $0.0288$ which is the strongest coupling
comparing to other ones shown in table \ref{tab:heffcpl}. However, if
the 3-phonon states are included this coupling vanishes
completely. This shows that a truncation of the computational basis
can not only cancel some indirect couplings \cite{obreschkow07} but
can also lead to the appearance of nonexistent ones. This increases the
number of optically active states in the resonance area and
affects the spectrum both qualitatively and quantitatively. 

Although the cancellation of an indirect interaction between certain 
2-phonon states in the presence of 3-phonon states is quite general 
and similar cancellation of the mediated interaction between other 
n-phonon states ($n>0$) can be observed it does not necessarily translate 
to the reduction of the size of the relevant polaronic subspace.
In general, such a decoupling is a quantitative effect and may depend
not only on the structure of the model, but also on the values of the 
couplings. For that reason the quasi-degenerate perturbation theory 
seems to be the right approach, which takes in to account the nested 
coupling structure \cite{obreschkow07}, as well as system dependent
quantitative effects.

The influence of $d$-shell and 3-phonons states on the diagonal
elements of the effective Hamiltonian is found to be less than $0.4\%$
and $2.5\%$ respectively.  Coupling with the two-phonon state
$|00,A^2B^{\bar1}\rangle$ is very weak and it is possible to discard
it from consideration. Such a reduction of the computational basis
does not produce any important effects.  The area of the second-order
resonant polarons can now be described by a Hamiltonian of dimension
5.

The results obtained with the effective Hamiltonian approach are nearly
the same as those obtained with full Hamiltonian
diagonalization. The largest shift between eigenenergies found with those
two methods in the range of the QD energy spacing presented in
Fig.~\ref{fig:porown-heff} is lower than 0.16~meV.

\subsection{Magnetopolaron resonance}\label{sec:HeffB}

The effective Hamiltonian treatment 
for a non-zero magnetic field
differs slightly from the previous case of
zero magnetic field, since the paramagnetic term brings the one-phonon
state $|0\bar1,A^2\rangle$ close to the second-order resonance. This
additional state couples to the purely electronic state and to certain
two-phonon states and, depending on the QD size, can be of major
importance. Nevertheless, the resonance area can be precisely
described using the quasi-degenerated perturbation theory even in the
case of relatively big QDs. When the previously mentioned one-phonon
state is too close to the energy of two LO phonons it simply needs to
be included as a member of the considered manifold. Since both samples
[Figs.~\ref{fig:sample3}(b), \ref{fig:sample3}(d)] consist of
relatively large QDs this one-phonon state has to be included in the
group of relevant states.

\begin{table}[tb]
\begin{displaymath}
\begin{array}{|c|c|c|} \hline \mathrm{matrix~elements} &
\multicolumn{1}{c|}{\mathrm{coupling~strength}} &d\mathrm{-shell~
influence}\\ \hline
\langle01|H_{\mathrm{eff}}|00,{A^2A^{\bar1}}\rangle & ~~1.786 & 7\% \\
\langle01|H_{\mathrm{eff}}|00,{A^0A^1}\rangle & ~~0.627 & 11\% \\
\langle01|H_{\mathrm{eff}}|00,{A^3 A^{\bar2}}\rangle & -0.157 &
100\%\\ \langle01|H_{\mathrm{eff}}|00,{A^0 B^1}\rangle & -0.143 &
100\%\\ \langle01|H_{\mathrm{eff}}|00,{A^2B^{\bar1}}\rangle & -0.056 &
100\%\\ \hline
\end{array}
\end{displaymath}
\caption{\label{tab:heffwField}Indirect effective
Hamiltonian couplings for case of magnetic field tuning. Values
obtained for electron excitation energy $\hbar
\omega_{0}=60~\mathrm{meV}$ in a 
magnetic field of $14.1~\mathrm{T}$, expressed in meV.}
\end{table} 

In the construction of the effective Hamiltonian, a set of 6
indirectly coupled states \{$|01\rangle$, $|00,A^0A^1\rangle$,
$|00,A^2A^{\bar 1}\rangle$, $|00,A^3A^{\bar 2}\rangle$,
$|00,A^0B^1\rangle$, $|00,A^2B^1\rangle$\} is chosen as a basis
(decoupling of certain 2-phonon states, discussed in the previous
section, is already taken into account). At this point, we are
interested in the strengths of indirect coupling mediated by 1-phonon
states, so we temporarily exclude the state $|0\bar1,A^2\rangle$ from
the manifold.  In this way, the influence of all intermediate 1-phonon
states on the second-order resonance can easily be determined. To
keep the model reasonably accurate in the absence of this state
Tab.~\ref{tab:heffwField} is 
calculated for a QD with a slightly larger characteristic energy
$\hbar \omega_0$, so that the energy separation between the relevant
manifold and the state $|0\bar1,A^2\rangle$ is also larger.  Indirect
coupling strengths for the resonance condition with the influence of
the $d$-shell are presented in Tab.~\ref{tab:heffwField}.
As one can see, in the case of the magnetopolaron
spectrum, the presence of the $d$-shell increases the coupling
strength significantly (rows 1-2) and
couples 3 additional 2-phonon states with the state $|01\rangle$
(rows 3-5). The character of the influence
of 3-phonon states is the same as that discussed in the previous
subsection.

\begin{figure}[tb]
\includegraphics[width=85mm]{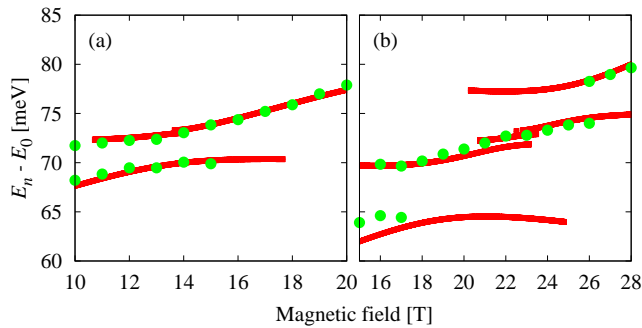}
\caption{\label{fig:heffB}(Color online) Polaron eigenenergies in the
resonance 
area. Dots: experimental results from
Refs.~\onlinecite{hameau99,hameau02}, 
lines: results 
obtained with the effective Hamiltonian approach. Samples and QD
material parameters as in Fig.~\ref{fig:sample3}.}
\end{figure}

As we pointed out earlier, in the case of relatively large QDs (when
the characteristic energy is lower than 60~meV), additional one-phonon
state in the effective Hamiltonian basis needs to be included. If this
is done then both methods: diagonalization of the full Hamiltonian and
quasi-degenerated perturbation theory approach produce nearly the same
results in the resonance area with differences in the obtained
eigenenergies and resonance widths smaller than $0.15~\mathrm{meV}$
and $0.01~\mathrm{meV}$, respectively. Comparison between the results
obtained with the effective Hamiltonian approach and the experimental
ones are presented in Fig.~\ref{fig:heffB}. Good agreement between the
theory and experiment can be observed.

\section{Conclusions} 

In this paper, we have theoretically studied the resonant features in
the spectrum 
of an electron confined in a self-assembled QD and interacting with LO
phonons. We have focused on the second-order resonance induced by the
indirect interaction between the 
first excited electronic shell ($p$-shell) and the electronic ground
state with two LO phonons. We have studied this second order resonant polaron
spectrum as a function of the dot size (energy level separation) and
external magnetic field. We have also calculated the absorption
spectra for an inhomogeneous ensemble of QDs and shown that polaronic
feature is clearly manifested in these spectra.  
Our results, compared to the existing experimental data, show that a
properly constructed model is able to  
quantitatively reproduce the observed polaron resonance without any
need for free or adjustable parameters describing the interaction 
between confined electrons and LO phonons, except for shape and size
parameters that can uniquely be extracted from the intraband absorption 
spectrum.

In order to get more insight into the structure of the polaron
spectrum in the resonance area, we have developed an effective
Hamiltonian approach based on a
quasi-degenerate perturbation theory. We have shown that by tracing
the structure of indirect couplings mediated by 1- and 3-phonon
states, a very small set of relevant basis states can be identified which
span the space of resonant polaron states. 

The presented results show that the spectrum of the coupled
electron--LO phonon system can be reliably modelled based on
the standard theories and computational techniques developed for
confined systems. Moreover, they demonstrate that this modeling may be
considerably simplified by applying perturbation theory methods,
without loosing the accuracy of the results. 

\appendix
\section{Influence of 4-phonon states on the ensemble
absorption} 
\label{appendixA} 
\begin{figure}[tb]
\begin{center}
\includegraphics[width=85mm]{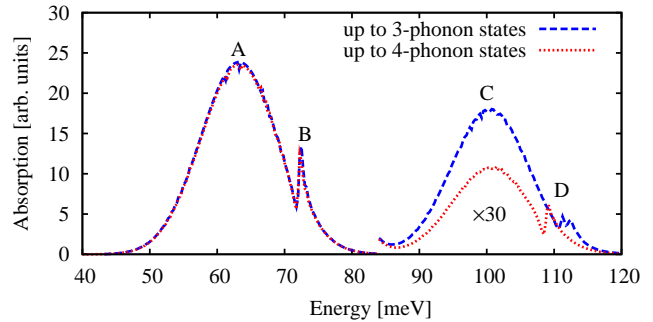}
\caption{(Color online) Absorption spectra on ensemble of isotropic
QDs obtained with and without 4-phonon states. Standard deviation
$6~\mathrm{meV}$, the mean value $\hbar\overline{\omega_0}= 62~\mev$.}
\label{fig:app}
\end{center}
\end{figure}

Since the 4-phonon states do not couple 0-
and 2- phonon states, their influence on the 2-phonon feature is
negligible. On the other hand, they are important if one considers
one-phonon replica of the second-order polarons. We compare here the
absorption spectra for two computational bases: one including only
states with up to 3 phonons and the other one with additional 4-phonon
states (Fig.~\ref{fig:app}). For the sake of simplicity, results were
obtained for the case of an isotropic QD without anharmonicity
corrections.

The influence of 4-phonon states on the main absorption peak (A), as
well as on 2-phonon resonance (B), is marginal. On the other hand,
4-phonon states are of major importance for one phonon replica
features (C,D).  Since those features appear as a result of the
interaction between 1-phonon and 3-phonon states, 4-phonon states have
similar influence on them as 3-phonon states had on the interaction
between 0- and 2- phonon states (decoupling of previously strongly
coupled states).  As a result of introducing additional interacting
states, a quantitative change is observed in the intensity of the phonon
replica (C) of the main absorption peak.  The most important change in
the ensemble absorption is related to the phonon replica of the
resonant feature (D).  If 4-phonon states are omitted this feature is
broader and consists of two peaks.  On the other hand, if we include
4-phonon states, the replica of the resonant feature consists of one
sharp peak, which is shifted towards lower energies. The energy shift
is mostly due to presence of directly coupled 4-phonon states with
energy $4\hbar \Omega$ which is higher than the energy of the feature D,
whereas the change in its shape is related to a reconstruction of the spectrum
in the area of the feature D, introduced by those additional states.


\end{document}